# Understanding the limits of remote focusing

SHARIKA MOHANAN,[1*] ALEXANDER D. CORBETT,[2,3]

[1]*School of Physics and Astronomy, University of Glasgow, G12 8QQ, UK.*
[2]*Department of Physics and Astronomy, University of Exeter, EX4 4QL, UK.*
[3]*Living Systems Institute, University of Exeter, EX4 4QL, UK.*

*\*sharika.mohanan@glasgow.ac.uk*

**Abstract:** It has previously been demonstrated in both simulation and experiment that well aligned remote focusing microscopes exhibit residual spherical aberration outside the focal plane. In this work, compensation of the residual spherical aberration is provided by the correction collar on the primary objective, controlled by a high precision stepper motor. A Shack-Hartmann wave front sensor is used to demonstrate the magnitude of the spherical aberration generated by the correction collar matches that predicted by an optical model of the objective lens. The limited impact of spherical aberration compensation on the diffraction limited range of the remote focusing system is described through a consideration of both on-axis and off-axis comatic and astigmatic aberrations, which are an inherent feature of remote focusing microscopes.

## 1. Introduction

### 1.1 Background

One of the current challenges in optical microscopy is the visualization of functional processes in organisms in their natural state. This requires the sample to be imaged unperturbed and with high fidelity in three dimensions. To accomplish this, a range of remote refocusing techniques allowing for agitation-free volumetric imaging have been implemented.

One such technique is remote focusing (RF) developed by Botcherby et al more than a decade ago [1]. One of the key advantages provided by RF is its ability to correct for high-NA defocus. Multifocus microscopes developed by Abrahamsson et al provides the same benefit but requires custom made diffractive elements that correct for a preselected FOV [2]. This leads to the second advantage of RF - its ability to instantaneously correct for aberrations within an extended volume. This contrasts with refocusing methods that use electro-tunable lenses, acousto-optic devices or adaptive optics where each defocused plane is corrected time sequentially [3], [4]. In RF, there is no requirement for optimizing the correction for each depth which leads to saving on the temporal bandwidth of the microscope. Light field microscopy (LFM) is another technique that can instantaneously image a volume. In LFM, a single camera frame captures the entire volume which leads to a trade-off between the axial and lateral resolution. In addition, reconstruction of the volume from a single image can result in processing artefacts [5].

In recent years, there has been a push for smarter microscopes that can meet the requirements for live sample imaging [6]. Due to the instantaneous correction of an extended volume, RF provides the basis to implement flexible scanning methodologies. This leads to sample adaptive scanning methods as demonstrated by Anselmi et al resulting in reduced data volume and decreased post-processing burden [7]. It also permits the construction of oblique plane microscopes (OPM) where conventional sample mounting procedures for light sheet microscopes can be implemented. Despite its advantages, RF systems have only been implemented in labs that have some amount of optics expertise. Which is to say that they are not a part of routine volumetric imaging of biological samples. The main limitation in building an RF system has to do with the alignment tolerances which becomes increasingly stringent as we move towards higher NA objectives [8].

**1.2 Working principle of RF system**

The schematic of an RF system built in the 'unfolded' geometry is shown in Fig. 1. Three microscopes (S1, S2 and S3) are aligned in series with the first two tube lenses (L1 and L2) forming the relay optics. The optical system from O1 to O2 is in telecentric alignment. The microscope S1 contains the imaging objective O1 which is closest to the sample being imaged and remains stationary. The microscope S2 demagnifies the intermediate image to form an aberration free remote volume around the focal plane of the refocusing objective O2. The reimaging objective, O3, scans this remote volume which is then imaged on to the detector. Another configuration of the RF system is the 'folded geometry' where O2 is reused as a reimaging objective by placing a mirror at the focal plane of O2. This mirror is axially translated to image the object at different depths. Due to the mirrors' lower inertia, the unfolded geometry allows for faster scan rates. However, in fluorescence mode, half of the signal is lost due to the presence of a polarizing beamsplitter placed immediately before O2.

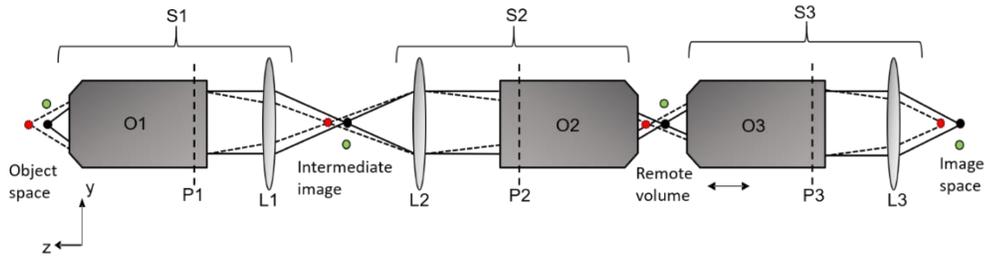

Fig. 1. An RF system in unfolded geometry with three microscopes S1, S2 and S3 in series. An intermediate image is formed with magnification $M_{S1}$. The remote volume has a uniform magnification of $M_{S1}M_{S2} = \frac{n_1}{n_2}$. The final image formed by S3 on the detector has a magnification of $M_{S1}M_{S2}M_{S3}$. The vertical dashed lines on the objectives signify the position of the exit pupil plane (P) for each objective.

Objectives designed for routine imaging of thin biological samples are optimized to follow the Abbe sine condition. As a zero condition, adherence results in the correction of off-axis coma which provides a diffraction-limited lateral (x-y) field of view (FOV). The RF system goes further by matching the Herschel condition which results in the cancellation of on-axis spherical aberration at defocused axial (z) positions. Botchery shows the method by which mapping the spatial frequencies in the pupil plane using the sine and Herschel condition leads to the disappearance of comatic terms and depth-dependent spherical aberration respectively (eqns. 9 to 12 in ref. [1]). In addition, the residual higher order spherical terms have been considered to define the axial limits of diffraction-limited imaging (eqn. 17 in ref. [1]).

In practice, it has been observed that when using high-NA objectives, RF provides only half of the theoretically determined diffraction-limited axial range. In the analytical formulism derived by Botcherby, there are assumptions made regarding the ideality of the objective and lenses used in the RF system. However, we need to consider the behavior of real systems to understand its practical limitations. In this paper we look at the effect of residual aberrations in the remote focusing system and how it limits the diffraction-limited 3D FOV from that predicted by theory. We first look at on-axis aberrations and the correction of residual spherical aberration. The amount of on-axis spherical aberration was measured using a Shack-Hartmann sensor and compensated for using the correction collar. Next, we investigated the effect of off-axis aberrations on the image quality at defocused positions.

## 2 Methods

### 2.1 Remote Focusing setup

The RF system presented in this paper was constructed to perform (1) widefield fluorescence imaging and (2) wavefront measurement (Fig. 2a and 2b respectively). The former was used to image sparse bead samples and the latter was used along with a Shack-Hartmann sensor for quantitative measurement of the phase distortion in the pupil plane of the final objective. The system was built in the unfolded geometry containing three microscopes in series. The first microscope was housed in an inverted Olympus IX73 stand. A 1.15 NA 40x water immersion objective (UAPON40XW340, Olympus) was used for O1. The correction collar on the objective was set to compensate for the spherical aberration introduced by the coverslip (#1.5 = 170 µm). The first tube lens forming the relay lens system (L1) is housed within the microscope stand and has a focal length of 180 mm. In this configuration, the distance between O1 and L1 is unknown. To keep the distance between the back focal plane of O1 and L1 fixed, O1 is placed in the lowermost position of the objective's axial travel range. This requires the sample to be brought towards O1's focal plane. This was done by mounting the sample along with a piezo translator, PT1 (Q-545.140, Physik Instrumente). The sample was illuminated using a Xenon arc lamp. A filter cube (GFP, Excitation:457 nm /Emission: 502 nm) was used to split the excitation and emission paths.

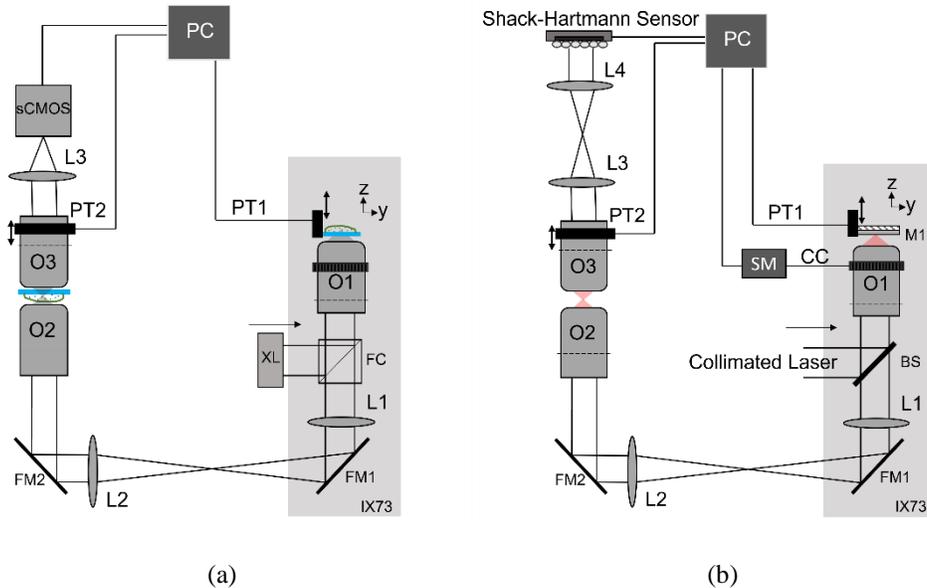

(a)  (b)

Fig. 2. RF system for (a) widefield fluorescence imaging and (b) pupil plane imaging with Shack-Hartmann sensor. BS – Beamsplitter, FM-Fold mirror, FC – Filter cube, PT-Piezo translator, SM -Stepper motor electronics, XL-Xenon lamp.

The rest of the RF system was aligned outside the microscope stand. The refocusing and reimaging objectives (O2 and O3) were 0.95 NA 40x dry objectives (UPLSAPO40X2,

Olympus). This required L2 to be a 135 mm focal length lens. The closest off-the-shelf lens has a focal length of 140 mm (G063235000, Qioptiq Ltd) was combined with a 4000 mm focal length lens to give an effective focal length of 135 ± 0.5 mm. O2 and O3 were aligned in a nose-to-nose configuration and glass coverslips (#1.5, 12 mm diameter) were carefully glued on to the front end of the objectives. O3 was translated using a piezo translator PT2 (P-725K085 PIFOC, Physik Instrumente) to refocus the system at different depths of the remote volume. The final tube lens, L3, has a focal length of 180 mm (#36-401, Edmund Optics). The images were captured using a sCMOS camera (Zyla 4.2, Andor Technology, Oxford Instruments).

### 2.1.1 Shack-Hartmann setup

We modified the widefield setup to measure the amount of spherical aberration in the remote focusing system using a Shack-Hartmann sensor. A 50:50 plate beamsplitter (Thorlabs, BSW10R) was mounted in the filter wheel of the Olympus IX73 to reflect the collimated laser light ($\lambda$ = 532 nm) towards O1. Mirror M1 was placed at the focal plane of O1. A thin layer of water was sandwiched between the mirror and a 170 µm coverslip with the space between the coverslip and the objective filled with a drop of water. M1 was defocused using the piezo translator PT1. The system was then refocused by translating O3 using PT2. The piezo translators were controlled using Micromanager software.

The Shack-Hartmann sensor was built using a microlens array (Thorlabs, MLA300-14AR-M) and a CMOS camera (iDS, UI-3240LE-M-GL). The camera sensor was placed at the focal plane of the lenslets (f = 14.6 mm). In the absence of aberrations, the wavefront is focused at the centre of each lenslet subaperture. For an aberrated wavefront, the spot is shifted proportional to the local slope of the wavefront over each subaperture. As the slope is the gradient of the wavefront over the subaperture, measuring the spot shifts provides a method to reconstruct the original wavefront. In reference [9], the list of the first 25 Zernike terms as used in this paper is reproduced in appendix 3D. The back focal plane of O3 was mapped onto the lenslet array of the Shack-Hartmann sensor using relay lenses L3 and L4. The final pupil diameter mapped on to the sensor is equal to the pupil diameter of O1 (10.35 mm) multiplied by the magnification of L1 and L2 (0.75x) and the magnification of L3 and L4 (0.6x).

### 2.2 Spherical Aberration Correction

Spherical aberration can be introduced in an RF system by (a) residual aberrations uncompensated by the refocusing and reimaging objectives and non-ideal alignment and (b) refractive index mismatch in the sample.

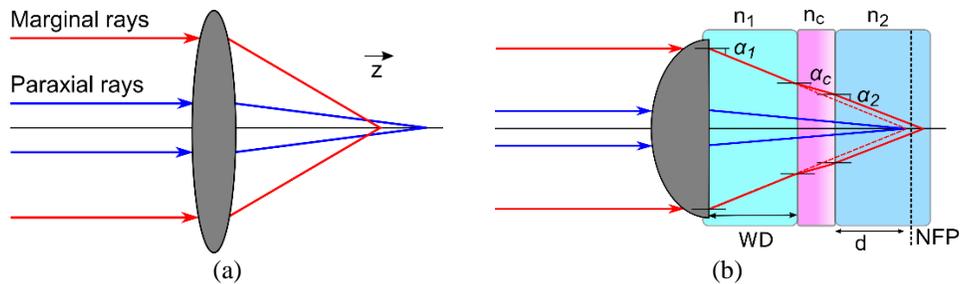

Fig. 3. (a) Lens-induced spherical aberration where the marginal rays are focused at a different axial position compared to the paraxial rays. (b) Specimen induced spherical aberration. The objective lens images into three different media - the immersion media having a refractive index of ($n_1$), the coverglass ($n_c$) and the sample ($n_2$). Immersion

objectives are corrected for the spherical aberration introduced by the coverslip. Additional spherical aberration is introduced if $n_2 \neq n_1$. $\alpha_1$, $\alpha_c$, and $\alpha_2$ are the angles of refraction for the converging marginal ray in the respective medium. In the presence of spherical aberration, the focal plane shifts to the nominal focal plane (NFP). WD is the working distance of the objective when d = 0.

Experimentally verified simulation of an ideal remote focusing system with an index-matched sample predicts the introduction of residual positive spherical aberration on either side of the focal plane [8]. A theoretical axial diffraction-limited range of 175 µm is predicted for a 1.15 NA 40x water immersion objective. Extension of this range via spherical aberration correction using the correction collar in the primary objective was investigated. O1 (UAPON40XW340, Olympus) has a correction range from 130 µm to 250 µm with major ticks every 40 µm and minor ticks every 20 µm. As repeatable and precise adjustment was required, the collar rotation was automated. To avoid confusion between the correction collar settings and the axial position in object space (both are in micrometres), the collar readings are prefixed by 'CC = '.

A stepper motor (Sanyo Denki, 103H5208-5240) was coupled to the correction collar using a timing belt (Fig. 3). The objective was fit with a 3D printed gear so that the timing belt could effectively grip the correction collar. The motor was controlled using an Arduino UNO microcontroller and an A4498 driver [10]. A novel addition is the use of an absolute encoder (Bourns, EMS22A) which was coupled to the shaft of the stepper motor to provide positional feedback. Feedback from the encoder allows us to set the positional limits at the ends of the rotation range of the collar. It also allows for the setting of a 'Home' position which is a reference position that the user can set (e.g.: 0.17 marking in the case of using a 170 µm coverslip). In our microscope, where the markings of the collar are positioned away from the user, the feedback gives us the position of the collar and therefore the compensation being applied at different depths. The code for the stepper motor control is made available here: https://github.com/sharika-mohanan/CorrectionCollarControl

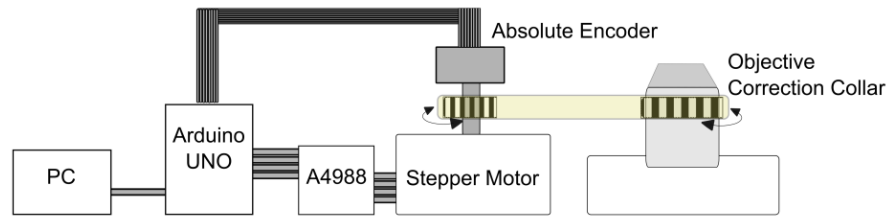

Fig. 3. Schematic showing the setup used to remotely set the position of the correction collar. The stepper motor is controlled using an Arduino UNO as the microcontroller and an A4988 driver. The gear on the motor shaft is coupled to the correction collar using a timing belt. The absolute encoder sends positional information back to the microcontroller. The user serially communicates with the microcontroller via the PC.

The collar rotates through an 120° arc to provide correction from CC = 130 µm to 250 µm. A single step of the motor allows for a 1.2 µm correction (change in coverslip thickness) to be applied. It was found that the spacing between the collar markings were not uniform across the 120° range and the rotation was calibrated to take this account. For the current experiments, the collar was rotated at low speeds to not disturb the sample. In addition to performing repeatable and accurate remote adjustment, the current implementation is cost effective with the entire setup costing less than £100.

We use the following formalism to predict the range of spherical aberration correction that can be provided by the correction collar [11]:

$$\text{OPD}_c = -n_1 \sin \alpha_1 \, d_t \left[ \sqrt{\frac{1}{\sin^2 \alpha_c} - \rho^2} - \sqrt{\frac{1}{\sin^2 \alpha_1} - \rho^2} \right] \quad (1)$$

$\text{OPD}_c$ is the optical path difference introduced by the correction collar for all rays across the pupil. $d_t$ is the difference between the correction collar position and the coverslip thickness. In our experiments, a 170 µm coverslip was used and if the collar was set to the CC = 150 µm position, $d_t$ = 20 µm. If an index matched immersion medium is used (in this case water), $\alpha_1$ is equal to the maximum acceptance angle of the objective (Fig. 3b) Therefore, the term $n_1 \sin \alpha_1$ is the NA of the objective. $\alpha_c$ is the angle of refraction of the marginal ray in the coverslip. $\alpha_c$ can be calculated using Snell's law where $n_c \sin \alpha_c = n_1 \sin \alpha_1$. For a glass coverslip, $n_c$ = 1.515. ρ is the normalised pupil radius. A negative sign is included as the correction collar introduces opposite amounts of aberration as introduced by the coverslip. The optical path difference across the pupil is then decomposed to individual Zernike terms to obtain the coefficient of aberration $C_p^q$ where $p$ is the axial order and $q$ is the azimuthal order of the expansion terms.

## 2.3 Bead samples

A standard method of measuring the PSF of a microscope is to image subresolution fluorescent beads which act as point sources. The beads were suspended in agarose gel prepared at a concentration of 1 % (w/v) to obtain a refractive index of ≈1.333 [12]. We used yellow-green microspheres of 100 nm diameter (F8803, ThermoFisher) with excitation at 505 nm and emission at 515 nm. The concentrated colloidal bead solution was diluted in ethanol (1:1000 µL). This was performed to achieve a sparse distribution of beads across the sample which then allows for volumetric imaging without the need for sectioning. The diluted bead solution was then mixed with the melted agarose gel (1:22 µL). The gel-bead mixture was allowed to set in a custom-made well chamber sealed using # 1.5 coverslips at both ends.

## 2.4 PSF Measurements

Image stacks of the bead sample were obtained by translating O3 every 0.2 µm using PT2 across a 400 µm range. The beads at every 5 µm depth was analysed using PSFj software [13]. With multiple beads present in each Field of View (FOV), the average axial FWHM of the PSF was calculated to get the resolution of the system at each depth. Using the definitions in [13], the theoretical FWHM for the current RF configuration can be calculated as:

$$\text{FWHM}_{x-y} = 0.84 \left( \frac{0.61 \, \lambda}{\text{NA}} \right) = 229 \text{ nm} \quad (2)$$

$$\text{FWHM}_z = 0.88 \left( \frac{2 \, n \, \lambda}{\text{NA}^2} \right) = 912 \text{ nm} \quad (3)$$

Here λ = 515 nm, n = 1.33 and NA = 1.15. It should be noted that the magnification of the remote volume is 1.33 x. Therefore, a translation of 0.2 µm in the remote volume space corresponds to a shift of 0.150 µm (= 0.2 µm/1.33) in sample space. Error bars indicate standard deviation of axial FWHM for beads taken within a specific lateral FOV.

## 3 Results and Discussion

### 3.1 On-axis aberrations

#### 3.1.1 Shack Hartmann wavefront aberration measurements

We first measured the dynamic range of spherical aberration compensation available to us using the correction collar. Mirror M1 was placed at the focal plane of O1 and the correction collar was rotated across its full range. Outside the CC=170 µm position, increasing amounts of spherical aberration was introduced in the system. The introduction of spherical aberration by the correction collar generates a small amount of defocus which is compensated by O3 to ensure $C_2^0 = 0$. From Fig. 5a, we observe that the experiment is consistent with the simulation predictions across the range shown. To measure the amount of residual spherical in the remote focusing system, M1 is displaced axially by a distance z and the system is refocused by translating O3 by 2 x 1.33 x z. The experimental remote focusing system has a diffraction-limited range of 90 µm (Fig. 5b). This is in line with the reduced amount of diffraction limited axial range seen in RF systems. When the correction collar is rotated incrementally towards the CC=130 µm position, negative spherical aberration cancels out the residual aberrations in the system in the defocused positions. For positive z, the maximum correction required at the 75 µm position was CC=154 µm. When high amounts of spherical aberration is present at defocused positions, such as in the −75 µm position, the collar was not able to not completely compensate even at CC=130 µm. We also observe an increased amount of second order spherical aberration in this region (Fig. 5d). After collar correction, we get an enhanced diffraction-limited range of 145 µm – an increase of ~60%.

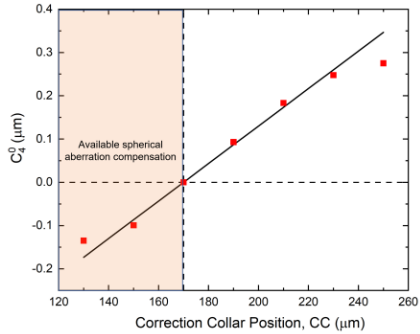

(a)

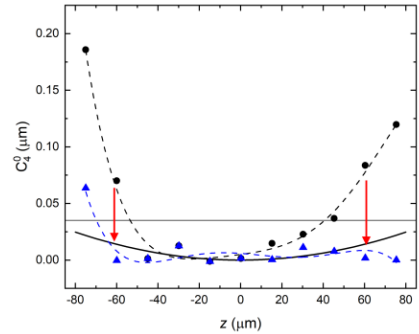

(b)

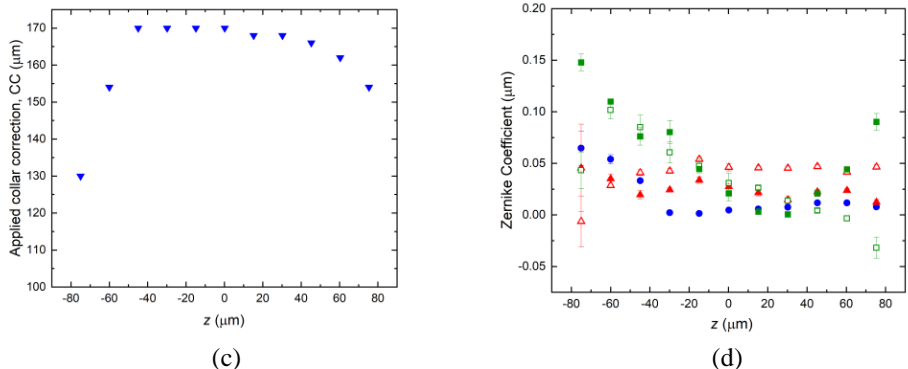

(c)                          (d)

Fig. 5: (a) Spherical aberration generated at different positions of the correction collar when using a 170 µm coverslip. Solid line is the simulation prediction, and the red squares are the experimental values. (b) Residual spherical aberration in an ideal remote focusing system. Solid black curve: Simulation prediction for ideal remote focusing system. Black circles: Experimental residual spherical aberration. Blue triangles: System corrected for spherical aberration using the correction collar. Black horizontal line indicates the threshold above which the Strehl ratio decreases below 0.8. (c) Position of the correction collar when correcting for residual spherical aberration in the RF system at different depths. (d) Residual aberration coefficient in the RF system. Solid blue circle: second order spherical aberration. Open red triangle: Y astigmatism. Solid red triangle: X astigmatism. Open green square: Y coma. Solid green square: X coma.

### 3.1.2 Bead PSF Measurements

The principle of SA correction is shown in Fig. 6a. Introduction of negative spherical aberration using the correction collar compensates for the residual aberration in the RF system. This brings the aberration coefficient $C_4^0 = 0$ at two positions symmetric to the focal plane. At these positions, the PSF should be diffraction limited. Axial resolution is determined from the axial FWHM of the fluorescent beads. Measurements were made at various depths of the sample volume averaged across half of the lateral FOV (148x148 µm). Fig. 6b shows that the Nominal Focal Plane (NFP) was shifted from the focal plane of O1 by +50 µm. This could be due to aberration balancing in the system. The presence of static spherical aberration also required the collar to be at 150 µm to reduce the aberrations over the entire range.

The anticipated variation in axial resolution is shown in Fig. 6b (red line). However, further introduction of negative spherical aberration did not improve the axial resolution at defocused positions (shown here for CC=130 µm in Fig. 6b). The system forms a typical 'W' curve where the introduction of negative spherical aberration reduces the resolution and increases the axial FWHM around the NFP of the system (Fig. 6a). However, it is expected to provide diffraction-limited imaging at two planes on either side of the NFP. The positions of these planes are expected to move outwards from the NFP as the amount of negative aberration introduced in the system increases. However, it seems the degradation of the axial resolution at defocused positions is not due to the presence of residual spherical aberration.

One possible explanation for this behaviour is the introduction of off-axis aberrations. This was verified by plotting the axial FWHM as a function of the distance from the optical axis at the NFP and a defocused position (-90 µm). In Fig. 6c, we show that near to the center of the lateral FOV, the increase in axial FWHM is negligible (< 100 nm) compared to its increase at the edges of the FOV.

**3.2 Off-axis aberrations**

As observed in Fig. 6, the resolution improvement expected from spherical aberration compensation in O1 did not materialize. We believe that this is due to the presence of uncompensated aberrations that mask the effect of the reduced spherical aberration. These uncompensated aberrations have two sources. The first is that when measuring the residual aberrations using the Shack-Hartmann sensor (Fig. 5d), odd aberrations such as coma and astigmatism are cancelled out due to the double-pass through O1 [14], thereby underestimating their contribution to the final PSF. The second reason is the presence of off-axis aberrations. The off-axis aberrations in an RF system cannot be quantified using the optical system used in Fig. 2b.

To further investigate how off-axis aberrations decrease in the diffraction limited lateral FOV with depth, we perform ray tracing simulation using Zemax (OpticStudio v15.5). As shown in Fig. 7a, two Olympus 1.4 NA 60x oil immersion objectives were used as O1 and O2. We have utilized the Zemax optical design for this objective as made available in reference [15]. The experimental system in Fig. 2 uses two doublet lenses for L1 and L2. As the optical design of these lenses are not available, we used two achromat doublets of 180 mm focal length (Thorlabs, AC508-180-A) in the Zemax model. The closest commercial Olympus objective is a 1.4 NA 60x apochromat objective 'PLAPON60XOSC2' which has a field number of 22 mm allowing for imaging across 366 µm lateral FOV.

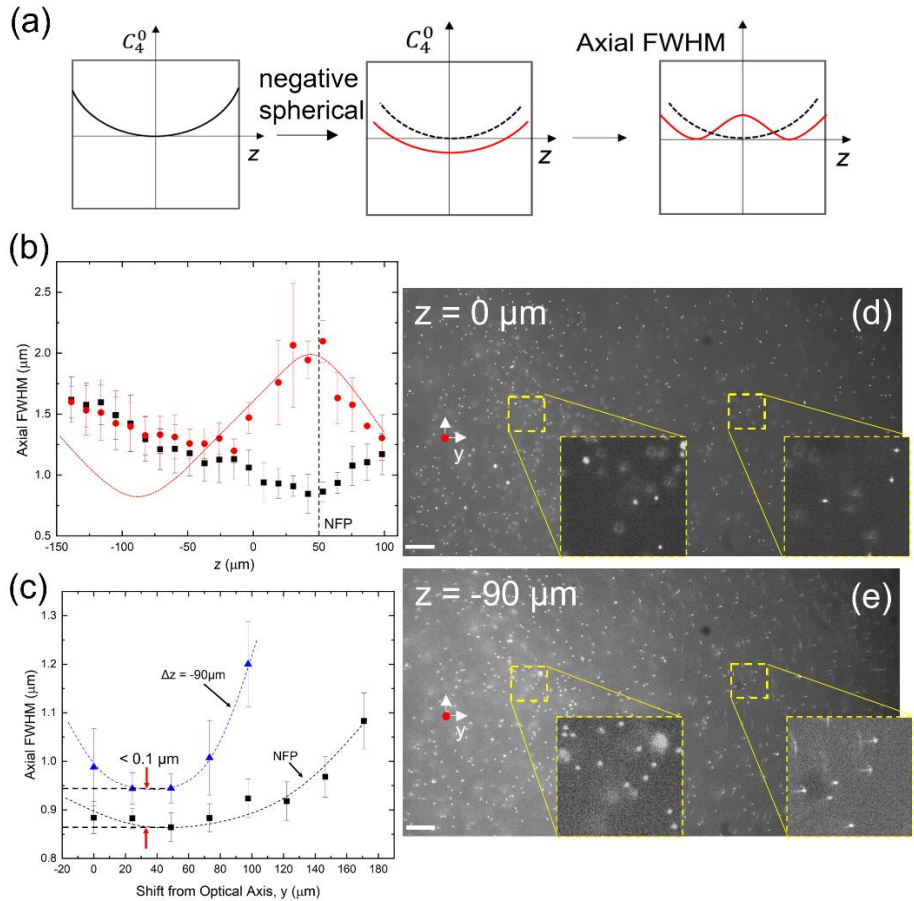

Fig. 6: (a) An ideal remote focusing system having positive spherical on either side of the focal plane requires the introduction of negative spherical to cancel the aberrations at defocused positions. It results in a 'W' shaped profile when measuring the axial FWHM. (b) Axial FWHM measured at defocused positions in the RF system. The focal plane is shifted from z = 0 to the NFP (z = 50 µm). Axial FWHM when CC=150 µm (black squares). Additional negative spherical is introduced by rotating the collar to the CC=130 µm position (red circles). The red dashed curve illustrates the expected trend. (c) Variation of axial FWHM with lateral FOV at the NFP (solid black squares) and at -90 µm (solid blue triangles). (d) Bead image at the focal plane. The red dot represents the position of the optical axis. (e) Bead image at -90 µm. The insets show the zoomed in lateral bead profiles with increasing aberration with distance from the optical axis. Scale bar 20 µm.

The simulation was performed for the wavelength $\lambda$= 587.5 nm. The system specifications showed the back focal plane of the objectives to be at -31.5 mm from the final lens element. The tube lenses were placed one focal length away form the back focal plane of the objectives. The change in paraxial magnification across the 80 µm axial range is less than 0.015 % ensuring telecentric alignment. The working distance (140 µm) of O1 was varied by changing the thickness of the immersion media layer resulting in change in the axial position of the object. Additionally, imaging at off-axis positions was studied by shifting the object ± 100 µm in x and y.

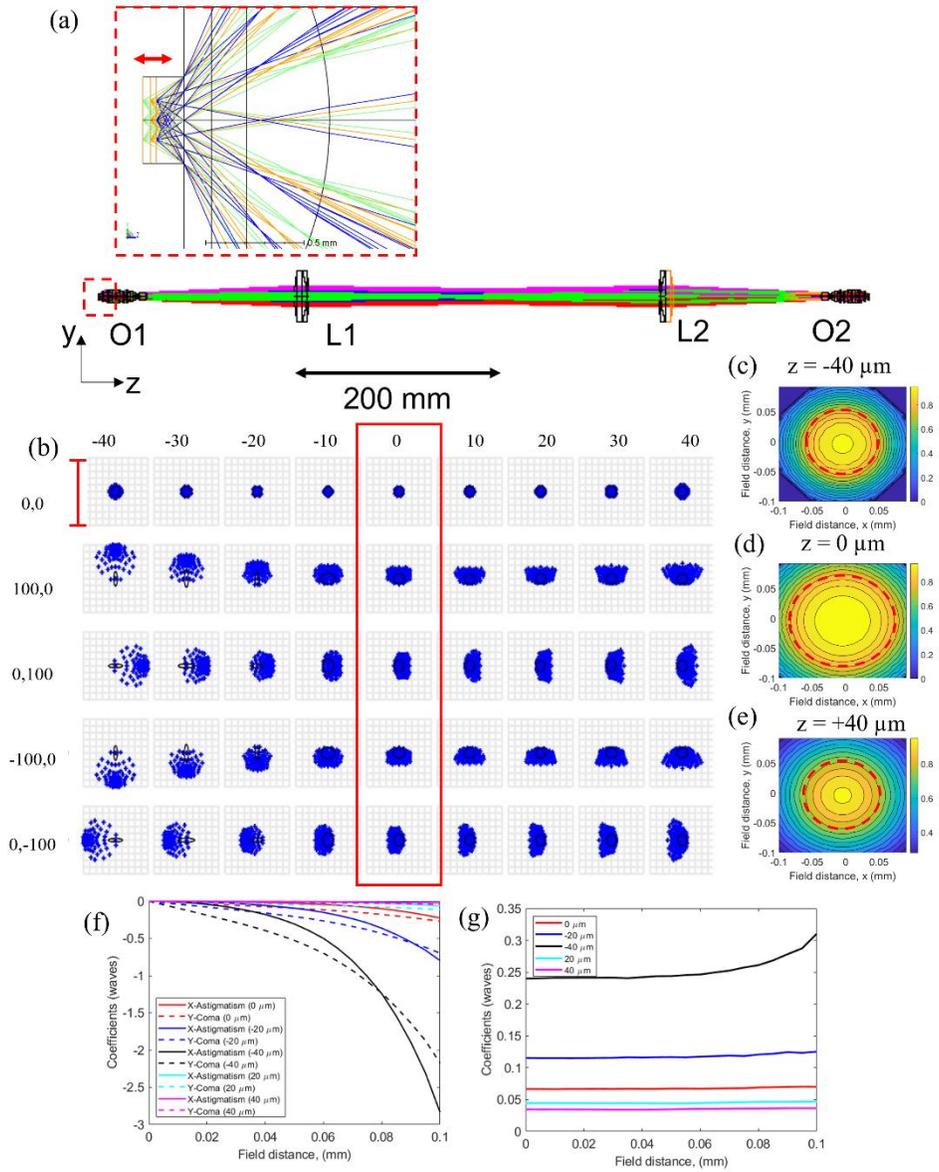

Fig. 7. (a) Optical layout of the RF system. Inset shows the zoomed in object space where the object is shifted in x,y and z. (b) Multi-configuration spot diagram showing the effect of off-axis imaging using an RF system. Each column corresponds to a 10 µm shift in axial position of the object with the red highlighted column corresponding to the native focal plane of the objective. The rows show the spot profile for different field positions (x,y). Units in micrometers. The black circle or ellipse in each graph corresponds to the Airy disk for on-axis and off-axis positions respectively. Scale bar = 5 µm. (c,d,e) Variation of Strehl ratio across the FOV. The red dashed line encircles the region within which the Strehl ratio is higher than 0.8. (f) Variation of X astigmatism and Y coma with field distance (y) at different defocused positions. (g) Variation of spherical aberrations with field distance.

The spot diagram for the various object positions is shown in Fig. 7b. The simulation is carried out across -40 µm to +40 µm from the focal plane in steps of 10 µm. It should be noted that the diffraction-limited axial range as predicted by Botcherby's theory for the 1.4NA oil immersion 60x objective is 80 µm. As expected, for on-axis position (x,y) = (0, 0), across the axial range, the system is diffraction-limited as the rays fall within the Airy disk (black ellipse) and the Strehl ratio is above 0.8 for a significant portion of the lateral FOV (Fig. 7d). However, when evaluating for off-axis positions for the same axial range, the system can no longer be considered diffraction-limited. This is reflected in the calculation of the Strehl ratio at positions $\pm$ 40 µm. The results from the simulation reflect the experimental observation regarding the decrease in the diffraction limited lateral FOV with increasing distance from the focal plane. The zemax simulation is made available here: https://github.com/sharika-mohanan/RF_zemax.

**3.3 Introduction of comatic aberration due to misalignment**

One additional effect observed in the RF system is the introduction of comatic aberration due to optical misalignment. Unlike uncompensated off-axis aberration discussed in the previous section – which increases with distance from the optical axis; coma introduced via misalignment is constant across the lateral FOV. Optical misalignment, in particular lateral misalignment between optical components can introduce a constant amount of coma across the FOV. It is also introduced when the coverslip on immersion objectives (especially for water and air lenses) is tilted [16].

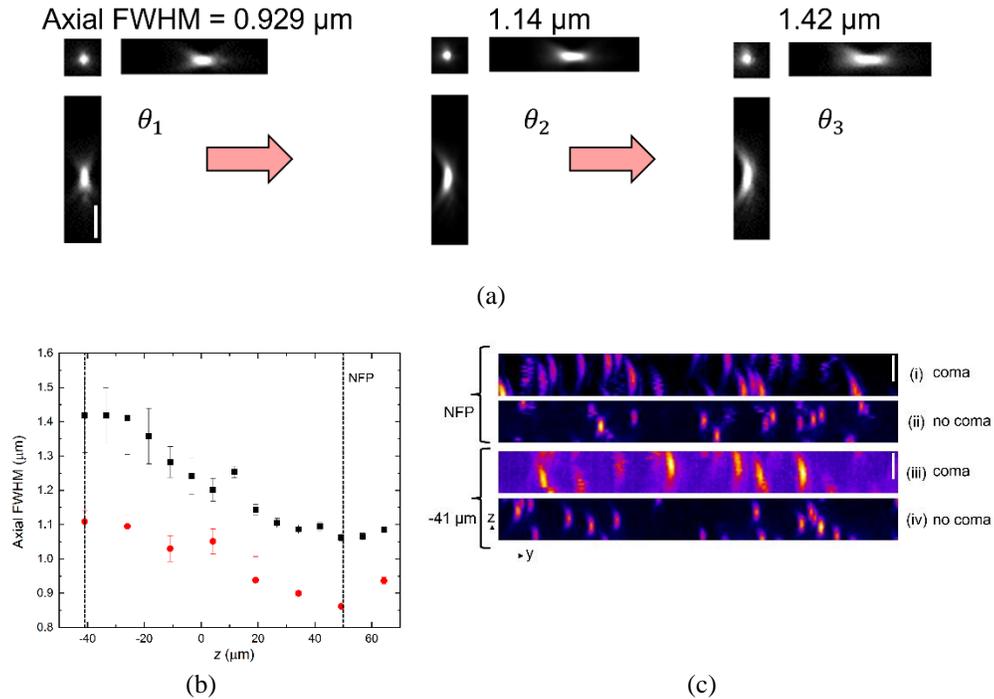

Fig. 8. (a) Coma introduced by tilting FM2, shown in Fig. 2, resulting increase in axial FWHM ($\theta_1 < \theta_2 < \theta_3$). Scale Bar = 2 µm (b) Axial FWHM due to coma introduced by coverslip tilt (black squares) and corrected for the same tilt (red circles). (c) yz projections of the axial FWHM at the z planes indicated by the vertical dashed lines in (b). (i)and (iii) correspond to the system with coma. (ii) and (iv) are corrected for coma. Scale Bar = 2 µm

We demonstrate the introduction of coma due to misalignment by tilting the beam entering O2 (Fig. 8). This is done by using the fold mirror FM2 in Fig. 2a. As the angular tilt of the mirror increases from $\theta_1$ to $\theta_3$, the amount of coma increases. This results in a characteristic curved axial profile when the axial projection is taken along the direction of the coma 'tail'. Looking at the individual bead profiles in Fig. 8a, it is evident that coma also reduces the axial resolution. This could be due to vignetting effects as the off-axis beam is clipped at the edges of the aperture stop.

The current optical configuration of the remote focusing system contains three objectives all requiring coverslips. The sample holder used with the piezo translator PT1, contains clamps to hold the sample in place. However, the clamping mechanism introduced sample tilt, therefore tilting the coverslip. The sample was imaged unclamped to remove coverslip induced coma. As seen in Fig. 8b, the axial FWHM for a clamped sample shows a significant decrease in axial resolution.

## 4. Conclusion

In this paper we have investigated the effect of residual aberrations in high-NA remote focusing systems. We have shown that correcting for on-axis spherical aberration can enhance the axial diffraction-limited range by using a correction collar on the immersion objective. However, when considering the entire lateral FOV, there is no significant improvement in image quality at defocused positions. One approach to correct for off-axis aberrations and enhance the diffraction-limited range is to implement adaptive optics correction [17]. As the RF system compensates for the bulk of spherical aberration generated at defocused positions, the deformable mirror used in the adaptive optics system would require a relatively small mechanical stroke to compensate for the residual aberrations. Such a microscope architecture would be complex to implement but provides the added flexibility to compensate for both specimen and system induced aberrations. It should also be noted here that adaptive optics compensation can limit temporal resolution either due to the requirement of optimizing an imaging metric or from SNR limitations.

The analytical formalism derived to explain the image formation in RF systems assumes ideal imaging for every point in the 3D FOV with a constant limiting angle '$\alpha$' for all rays entering the objective. In other words, it assumes that the physical aperture within the objective lens that limits the rays is the same for all defocused positions. This brings us to the crux of the problem in trying to predict the behaviour of RF systems. As the design of objective lenses are not freely available, it is difficult to model their response to imaging outside its optimum design conditions. It is known that at the edges of the lateral field on the focal plane, vignetting effects can reduce the throughput and therefore the effective NA at these positions [15], [18]. This could potentially be an added limitation if the vignetting effects increase away from the focal plane. To truly optimize RF imaging across the 3D volume, we must start by redesigning the objective using a different set of optimization criteria that provides equal weighing of points inside and outside the focal plane at both on- and off-axis positions. This is in line with recent development of microscope objectives that have been built to suit a specific application [19], [20]. Such an objective could lead to a decrease in the complexity of alignment and wider application in routine 3D volumetric imaging.

**Acknowledgments**

We would like to thank Prof. Christian Soeller and Prof. Martin Booth for helpful discussions while working on this project.

We would like to thank Prof. Christian Soeller and Prof. Martin Booth for helpful discussions while working on this project.